\documentclass[10pt,conference]{IEEEtran}
\usepackage{xspace}
\usepackage{mdframed}
\usepackage{color, colortbl}
\usepackage{array}
\usepackage{hyperref}
\usepackage{graphicx}
\usepackage{multirow}
\usepackage{soul}
\usepackage{cleveref}
\usepackage{multirow}
\usepackage{booktabs}
\usepackage{graphicx}
\usepackage[table,xcdraw]{xcolor}

\usepackage{fancybox}
\newcommand{\hypobox}[1]{

        \begin{center}\noindent\thicklines\setlength{\fboxsep}{6pt}\cornersize{0.2}\ovalbox{

                \begin{minipage}{3.0in}

                        \textit{#1}

                \end{minipage}} 

        \end{center}} 

\mdfdefinestyle{MyFrame}{
 outerlinewidth=0pt,
 skipabove=0pt,
 skipbelow=0pt,
 innertopmargin=5pt,
 innerbottommargin=0pt,
 linewidth=0pt,
 topline=false,
 rightline=false,
 leftline=false,
 innerrightmargin=4pt,
 innerleftmargin=4pt,
 nobreak=false}

\newcommand{\GH}{{\sc GitHub}\xspace}
\newcommand{\DP}{{\sc Devpost}\xspace}
\newcommand{\WOC}{{\sc World of Code}\xspace}
\newcommand{\RQ}[2]{
\begin{mdframed}[style=MyFrame]\noindent
	\textbf{RQ}$_{#1}$.~\emph{#2}
\end{mdframed}
}

\makeatletter
\newcommand{\linebreakand}{%
  \end{@IEEEauthorhalign}
  \hfill\mbox{}\par
  \mbox{}\hfill\begin{@IEEEauthorhalign}
}
\makeatother

\definecolor{Gray}{gray}{0.9}

\newcolumntype{L}[1]{>{\raggedright\let\newline\\\arraybackslash\hspace{0pt}}m{#1}}

\usepackage{ifthen}
\usepackage{amssymb}
\DeclareOldFontCommand{\sf}{\normalfont\sffamily}{\mathsf}
\newboolean{showcomments}
\setboolean{showcomments}{true}
\ifthenelse{\boolean{showcomments}}
  {\newcommand{\nb}[2]{
    \fcolorbox{Gray}{yellow}{\bfseries\sffamily\scriptsize#1}
    {\sf\small$\blacktriangleright$\textit{#2}$\blacktriangleleft$}
   }
   
  }
  {\newcommand{\nb}[2]{}
   
  }

\begin{document}
\bstctlcite{IEEEexample:BSTcontrol}

\title{Tracking Hackathon Code Creation and Reuse}

\author{\IEEEauthorblockN{Ahmed Imam}
\IEEEauthorblockA{University of Tartu\\
Estonia\\
ahmed.imam.mahmoud@ut.ee}
\and
\IEEEauthorblockN{Tapajit Dey}
\IEEEauthorblockA{Lero---the Irish Software Research\\
Centre, University of Limerick\\
Limerick, Ireland\\
tapajit.dey@lero.ie}}

\maketitle
\thispagestyle{plain}
\pagestyle{plain}

\begin{abstract}
Background: Hackathons have become popular events for teams to collaborate on projects and develop software prototypes. Most existing research focuses on activities during an event with limited attention to the evolution of the code brought to or created during a hackathon.
Aim: We aim to understand the evolution of hackathon-related code, specifically, how much hackathon teams rely on pre-existing code or how much new code they develop during a hackathon. Moreover, we aim to understand if and where that code gets reused.
Method: We collected information about 22,183 hackathon projects from \DP -- a hackathon database -- and obtained related code (blobs), authors, and project characteristics from the \WOC. We investigated if code blobs in hackathon projects were created before, during, or after an event by identifying the original blob creation date and author, and also checked if the original author was a hackathon project member. We tracked code reuse by first identifying all commits containing blobs created during an event before determining all projects that contain those commits.
Result: While only approximately 9.14\% of the code blobs are created during hackathons, this amount is still significant considering time and member constraints of such events. Approximately a third of these code blobs get reused in other projects.
Conclusion: Our study demonstrates to what extent pre-existing code is used and new code is created during a hackathon and how much of it is reused elsewhere afterwards. Our findings help to better understand code reuse as a phenomenon and the role of hackathons in this context and can serve as a starting point for further studies in this area. 
\end{abstract}




\section{Introduction}
\label{sec:intro}
Hackathons are time-bounded events during which individuals form -- often ad-hoc -- teams and engage in intensive collaboration to complete a project that is of interest to them \cite{pe2019designing}. Most hackathon projects focus on creating a prototype that can be presented at the end of an event~\cite{medina2020what}. This prototype often takes the form of a piece of software. The creation of software code can, in fact, be considered as one of the main motivations for organizers to run a hackathon event. Scientific and open source communities, in particular, organize such events with the aim to expand their code base~\cite{pe2019understanding,stoltzfus2017community}. 
It thus appears surprising that the evolution of the code used and developed during a hackathon has not been studied yet, as revealed by a review of existing literature. In our paper, we aim to address this knowledge gap by studying 22,183 hackathon projects, identified using\DP, by leveraging \WOC, a dataset of almost all open source projects.

\textbf{Complete results of an extension of this hackathon project is available at}~\cite{imam2021secret}, and \textbf{the replication package} for our study is available at~\cite{repPackage}.

\section{Research questions}

In order to address the knowledge gap mentioned earlier, in this hackathon project we aimed to study the evolution of the code used and created by the hackathon team members from two main perspectives. First, we studied where the code \textit{originates}: While teams will certainly develop original code during the hackathon, it can be expected that they will also utilize existing (open source) code as well as code that they might have created themselves prior to the event, so our first research question that addresses the topic:

\RQ{1}{ Where does hackathon code come from?}
\vspace{-8pt}

\noindent In particular, we focused on the sub-questions:

\RQ{1a}{When was the code created?} \vspace{-5pt}
\RQ{1b}{Who were the original creators of the code?}\vspace{-8pt}

Second, to understand the impact of hackathon code, i.e. code created during a hackathon event by the hackathon team in the hackathon project repository, on the wider software development community, we aimed to study whether and how it \textit{propagates} after the event has ended. As noted in~\cref{sec:intro}, existing studies do not address the question of whether and where hackathon code gets reused after an event has ended. In fact, hackathons are widely considered as ``one-off'' events by many. Knowing the answer to this question, thus, would be crucial for understanding the impact of hackathons on the larger open source community. This leads us to also asking the following second research question:
\RQ{2}{What happens to hackathon code after the event?}
\vspace{-8pt}

\section{Methodology and Results}
\label{sec:method}

To address our research questions, we conducted an archival analysis of the source code utilized and developed in the context of 22,183 hackathon projects that were listed in the hackathon database \DP\footnote{https://devpost.com/}. To track the origin of the code that was used and developed by each hackathon project and study its reuse after an event has ended, we leveraged the open-source database \WOC \cite{ma2019world, ma2020world}, the primary focus of this hackathon event, which allowed us to track the origin of hackathon code and code reuse across almost all open source repositories. In our study, we focused on blob-level code reuse.

\subsection{Data Sources}

\DP is a popular hackathon database that is used by corporations, universities, civic engagement groups and others to advertise events and attract participants. It contains data about hackathons including hackathon locations, dates, prizes and information about teams and their projects including the project's \GH repositories. It was our primary source for identifying the ``hackathons''.

However, \DP doesn't have all the information we need for answering our research questions, so we leveraged the \WOC dataset for gathering additional information about the projects, authors, commits, and code blobs. 

\vspace{-5pt}
\subsection{Data Collection and Analysis}

\begin{figure}[t]
    \centering
    \includegraphics[width=\linewidth]{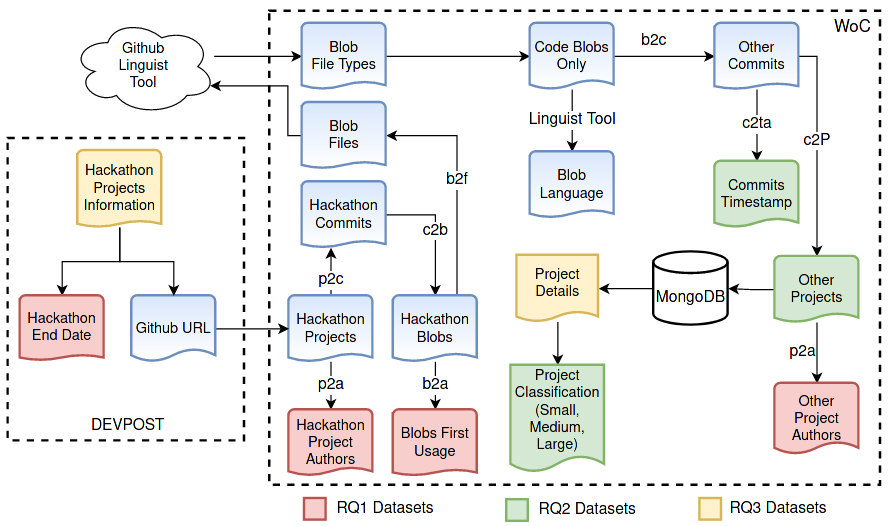}
    \caption{Data Collection Workflow: Highlighting the different data sources used and the process of gathering the required information from them, and the data used in answering our research questions}
    \label{fig:DataCollection}
    \vspace{-10pt}
\end{figure}

Here we describe how we collected the data required for answering our research questions, along with details of all the filtering we introduced. An overview of the approach is shown in~\cref{fig:DataCollection}, which also highlights the different data sources and what data was used for answering each research question.

\subsubsection{Steps to answer RQ1}
Our starting point was a list of hackathon projects from \DP and using the \GH URLs for the projects, we mapped them to the project names in \WOC, which are formatted as \texttt{GitHubUserName\_RepoName}. We started by collecting all the commits for these projects using \textit{p2c} map, then we collected all the blobs in those commits using \textit{c2b} map. 
We were also interested in identifying the authors so we collected the authors for the hackathon projects using \textit{p2a} map and the first author who introduced each blob along with the timestamp when the blob was first used using \textit{b2a} map.

It is common in projects to not only have code files but also other file types like images, data, markup, and prose files, so we collected the file information of each blob using \textit{b2f} map which was important to identify non-code blobs and filter them out during our analysis. We utilized the \textit{linguist} tool from \GH to find out the file types of the blobs and we filtered only code blobs.

In order to answer our first research question, we started by understanding \textbf{``when''} the blob is first used with compared with the hackathon event start and end dates. The \DP dataset doesn't include the start date, so we derived the start date from the end date by assuming the duration of hackathon events of 72 hours which appears reasonable since hackathons are commonly hosted over a period of 48 which are often distributed over three days~\cite{Nolte2020HowTO,cobham2017appfest2}. We then compared the first timestamp of each blob with the hackathon event start and end dates and we classified the blobs based on time to before, during, and after the hackathon event.

Since we are also interested in \textbf{``who''} are the original creators of the blobs and their connection with the hackathon projects in question, we checked if the original author of a hackathon code blob was part of the hackathon team or not. We also wanted to understand if any of the hackathon project members were part of the other projects where the blobs were first introduced (we call the code creators as ``co-contributors'' in such cases). Since we have the first commit which introduced this blob from the \textit{b2a} map, we collected the projects for these commits using \textit{c2P} map and then the author list of these projects using \textit{p2a} map.  We used the approach outlined by~\cite{fry2020idres} for author ID disambiguation to merge all of the different IDs belonging to one developer together, which is a common occurrence, as discussed in~\cite{Dey2020RepresentationOD}.

\begin{figure}[t]
\centering
\vspace{-10pt}
\includegraphics[width=\linewidth]{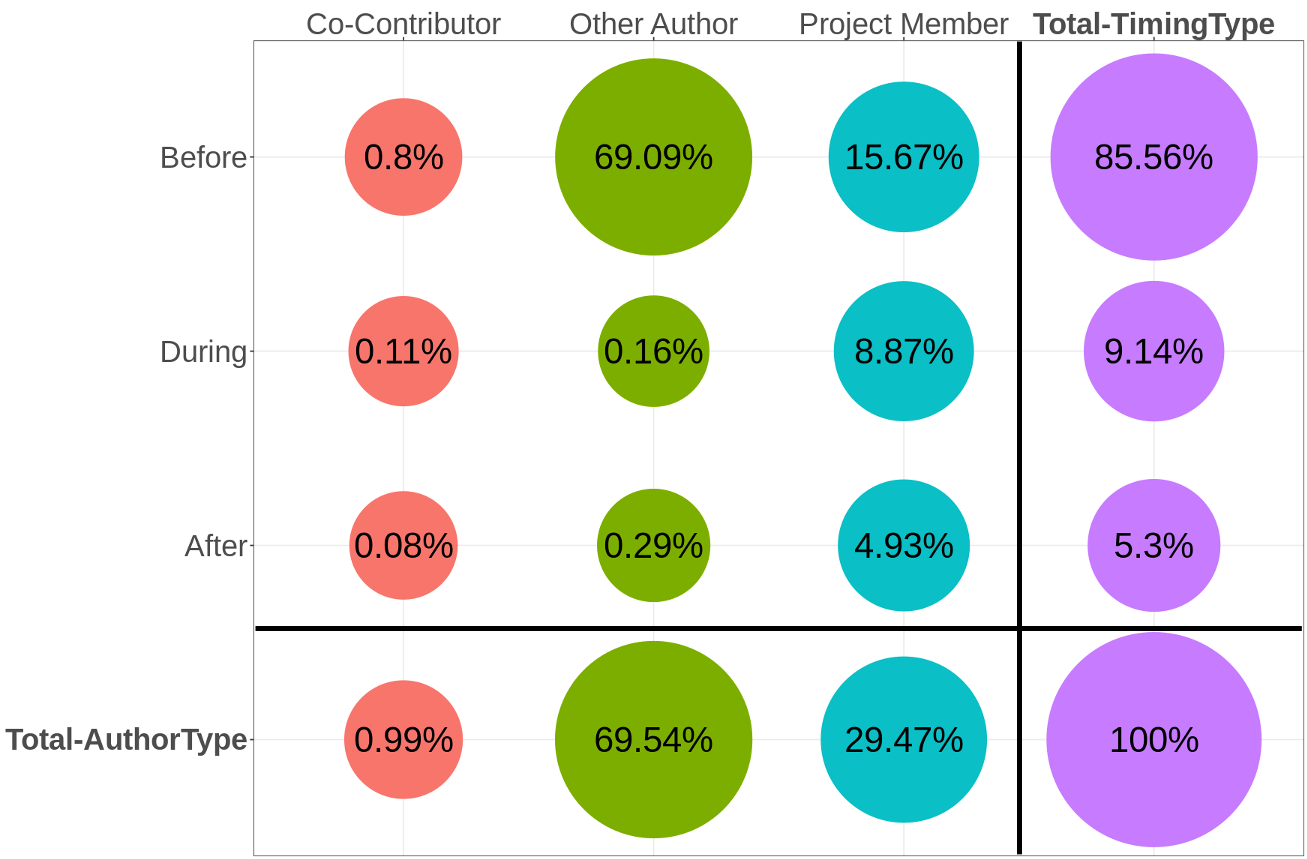}%
\vspace{-10pt}
\caption{Plot depicting what percentage of the hackathon code was created by whom and when}
\label{fig:rq1}
\vspace{-10pt}
\end{figure}

Figure \ref{fig:rq1} shows the results of our analysis for RQ1, highlighting that 85.56\% of the code (in terms of the no. of blobs) in the hackathon project repositories is created before the hackathons, with around 9.14\% of the code being created during the events (which is significant considering the duration and team member constraints of the hackathons). Moreover,  The members of the hackathon teams created around 29.47\% of the code blobs, while 69.54\% of the code blobs are created by developers outside the team (mostly authors of some project/package/framework used by the team).

\hypobox{\textbf{Origin of the Hackathon Code (RQ1)}: Hackathon projects often reuse code in terms of some package/framework. Teams also tend to reuse their own code. Most of the code created during or after the event is created by the hackathon team members.}

\subsubsection{Steps to answer RQ2}
Here our goal was to understand how the hackathon-generated code gets reused after the hackathon event, so our starting point was the result from the RQ1 analysis since we used that as a base for filtering and answering RQ2. We applied a filter to blobs that satisfy two conditions: (a) Blobs are created during the hackathon event and (b) Blobs are created by hackathon project team members. Once these blobs are identified, we start collecting the commits that use these blobs using \textit{b2c} map, and we collected the commit timestamps using \textit{c2ta} map. We also used the project information dataset from a Mongo Database associated with \WOC to identify the project size using two variables \textit{(numAuthors, numStars)} which are indications of project size and popularity and were found to have a low correlation (Spearman Correlation: 0.26). We used \textit{Hartemink's pairwise mutual information-based discretization method}~\cite{hartemink2001principled}, which was applied to a dataset with log-transformed values of the number of stars and developers for the projects, to classify the projects into three categories: Small, Medium, and Large. 89.2\% of the projects that reused the hackathon code blobs were classified as \textit{Small}, 8.5\% were \textit{Medium}, and 2.3\% were classified as \textit{Large} projects.


\hypobox{\textbf{Hackathon code reuse (RQ2):} Around 28.8\% of hackathon code blobs got reused in other projects, with 57.73\% of the code being used in \textit{Small} projects, 32.85\% in \textit{Medium} projects, and 9.42\% in \textit{Large} projects. Most of the reused blobs were related to web/mobile apps/frameworks. The temporal dynamics of code reuse show a clear trend of it reducing over time, and then saturating to a stable value.}


\section{Future work}
\label{sec:future}
There are several ways to extend this research, e.g. considering code clones/snippets while looking for code reuse (e.g. by looking at the associated CTAG tokens - a dataset available in \WOC), identifying other factors that affect code reuse, including code quality~\cite{dey2018usageQuality,Dey2020qualityEMSE}, project popularity~\cite{dey2018dependency,dey2019patterns}, the type of Open Source license used, etc. Looking deeper into the code created during the hackathons, it might also be interesting to see to what extent the teams use bots~\cite{dey2020botdetection,dey2020botse} which might aid in the understanding of hackathon code reuse as well.
We hope that further studies will explore these and other related topics, and give us a clearer understanding of the impact of hackathons and code reuse.




\bibliographystyle{IEEEtran}
\bibliography{references}

\end{document}